\documentclass[12pt]{article}
\usepackage{graphicx} % Required for inserting images
\usepackage{bm}
\usepackage{authblk}
\usepackage{amsmath}
\usepackage{graphicx}
\usepackage{xcolor}
\usepackage{abstract}
\usepackage{soul}

\usepackage[letterpaper]{geometry}
\makeatletter
\renewcommand{\maketitle}{\bgroup\setlength{\parindent}{0pt}
\begin{flushleft}
  \textbf{\@title}

  \@author
\end{flushleft}\egroup
}
\let\saved@includegraphics\includegraphics
\AtBeginDocument{\let\includegraphics\saved@includegraphics}
\renewenvironment*{figure}{\@float{figure}}{\end@float}

\makeatother
\usepackage[labelfont=bf]{caption}

\usepackage[backend=biber,style=nature,sorting=none,maxnames=99,minnames=1]{biblatex}
\renewcommand*{\supercite}[1]{\textsuperscript{\cite{#1}}}
\DeclareFieldFormat{labelnumber}{#1} 
\AtEveryBibitem{%
  \ifentrytype{incollection}{%
    \DeclareFieldFormat{title}{#1}%
  }{}%
}

\DeclareBibliographyDriver{incollection}{%
  \printnames{author}%
  \adddotspace%
  \printfield{title}%
  \newunit
  \printfield{booktitle}%
  \newunit
  \printfield{volume}%
  \newunit
  \printfield{number}%
  \newunit
  \printfield{pages}%
  \newunit
  \printfield{year}%
}
\title{Turn-on of Current-Induced Spin Torque upon Noncollinear Antiferromagnetic Ordering in Delafossite PdCrO$_2$ %\let\thefootnote\relax\footnote{
}
\author[1$\dagger\ast$]{Xiaoxi Huang}
\author[2$\dagger$]{Qi Song}
\author[3,4]{Gautam Gurung}
\author[1]{Daniel A. Pharis}
\author[1]{Thow Min Jerald Cham}
\author[2]{Yulan Chen}
\author[1]{Rakshit Jain}
\author[1]{Maciej Olszewski}
\author[2]{Yufan Feng}
\author[6]{Amal El-Ghazaly}
\author[4]{Evgeny Y. Tsymbal}
\author[2,5,7]{Darrell G. Schlom}
\author[1,5,$\ast$]{Daniel C. Ralph}

\affil[1]{Department of Physics, Cornell University, Ithaca, NY 14853, USA.}
\affil[2]{Department of Materials Science and Engineering, Cornell University, Ithaca, NY 14853, USA.}
\affil[3]{Trinity College, University of Oxford, Oxford, OX1 3BH, United Kingdom.}
\affil[4]{Department of Physics and Astronomy \& Nebraska Center for Materials and Nanoscience, University of Nebraska, Lincoln, NE 68588, USA. }

\affil[5]{Kavli Institute at Cornell for Nanoscale Science, Ithaca, NY 14853, USA.}
\affil[6]{Electrical and Computer Engineering, Cornell University, Ithaca, NY 14853, USA.}
\affil[7]{Leibniz-Institut fur Kristallzüchtung, Max-Born-Str. 2, 12489 Berlin, Germany.}
\affil[$\dagger$]{These authors contributed equally to this work.}
\affil[$\ast$]{Corresponding authors: xh384@cornell.edu, dcr14@cornell.edu}
\date{September 2023}
\begin{document}
    \maketitle
    \vspace{10mm}
    \noindent
    \textbf{Abstract}\\
    \textbf{We report measurements of the current-induced spin torque produced by the delafossite antiferromagnet PdCrO$_2$ and acting on an adjacent ferromagnetic permalloy layer.  The spin torque increases strongly as the temperature is reduced through the N\'eel temperature, when the PdCrO$_2$ transitions from a paramagnetic phase to a noncollinear antiferromagnetic state. 
    %The spin torque is negligible for the paramagnetic phase of PdCrO$_2$, but it increases strongly as the temperature is reduced through the N\'eel temperature.  
    This result is qualitatively consistent with density functional theory calculations regarding how spin-current generation changes upon antiferromagnetic ordering in PdCrO$_2$.  
    %However the low-temperature spin torque efficiency per unit current density and torque efficiency per unit electric field are merely modest ($\xi_{SH} = 0.027$, $\sigma_{SH} = 0.68 \times 10^4$ $\Omega^{-1}$m$^{-1}$). 
    %but because of the low resistivity of PdCrO$_2$ ($\rho$ = 45 $\mu\Omega$cm at 4.2 K) the torque efficiency per unit electric field is quite high ($\sigma_{SH} > 4 \times 10^6$ $\Omega^{-1}$m$^{-1}$.)  {\color{blue}[I'll check numbers here. The resistivity is actually not all that low, since Pt by itself has $\rho$ = 20 $\mu\Omega$cm.]}
    }\\
    
    \noindent
    \textbf{Introduction}\\
    Spin torques exerted by spin currents which flow transverse to an applied charge current offer a strategy for improving the efficiency of magnetization manipulation for spintronic applications. This has triggered an extensive search for materials which can serve as the most efficient sources of such spin currents. Conventionally, a prerequisite for current-induced torque generation has been strong spin-orbit coupling (SOC) within either a non-topological\supercite{mihai2010current,miron2011perpendicular,liu2011spin,liu2012spin,garello2013symmetry,nan2019anisotropic,huang2021novel} or topological electronic band structure,\supercite{mellnik2014spin,fan2014magnetization,fan2016electric} which limits the potential material candidates.  It is now understood that both non-collinear\supercite{vzelezny2017spin, zhang2018spin} and collinear\supercite{gonzalez2021efficient} antiferromagnetic orderings allow for the generation of transverse-flowing spin currents even in the absence of SOC, and this has spawned several experimental studies of antiferromagnets as sources of spin torque.\supercite{zhang2016giant,vzelezny2017spin,zhang2017strong,zhang2018spin,kimata2019magnetic,holanda2020magnetic,nan2020controlling,chen2021observation,bose2022tilted,gonzalez2021efficient,hu2022efficient} Nevertheless, it remains challenging to unambiguously identify a system where spin current is produced primarily by antiferromagnetic ordering, owing to the presence of SOC in most antiferromagnets.
    
\begin{figure}[!hb]
    	\centering
        \includegraphics[width=1\textwidth]{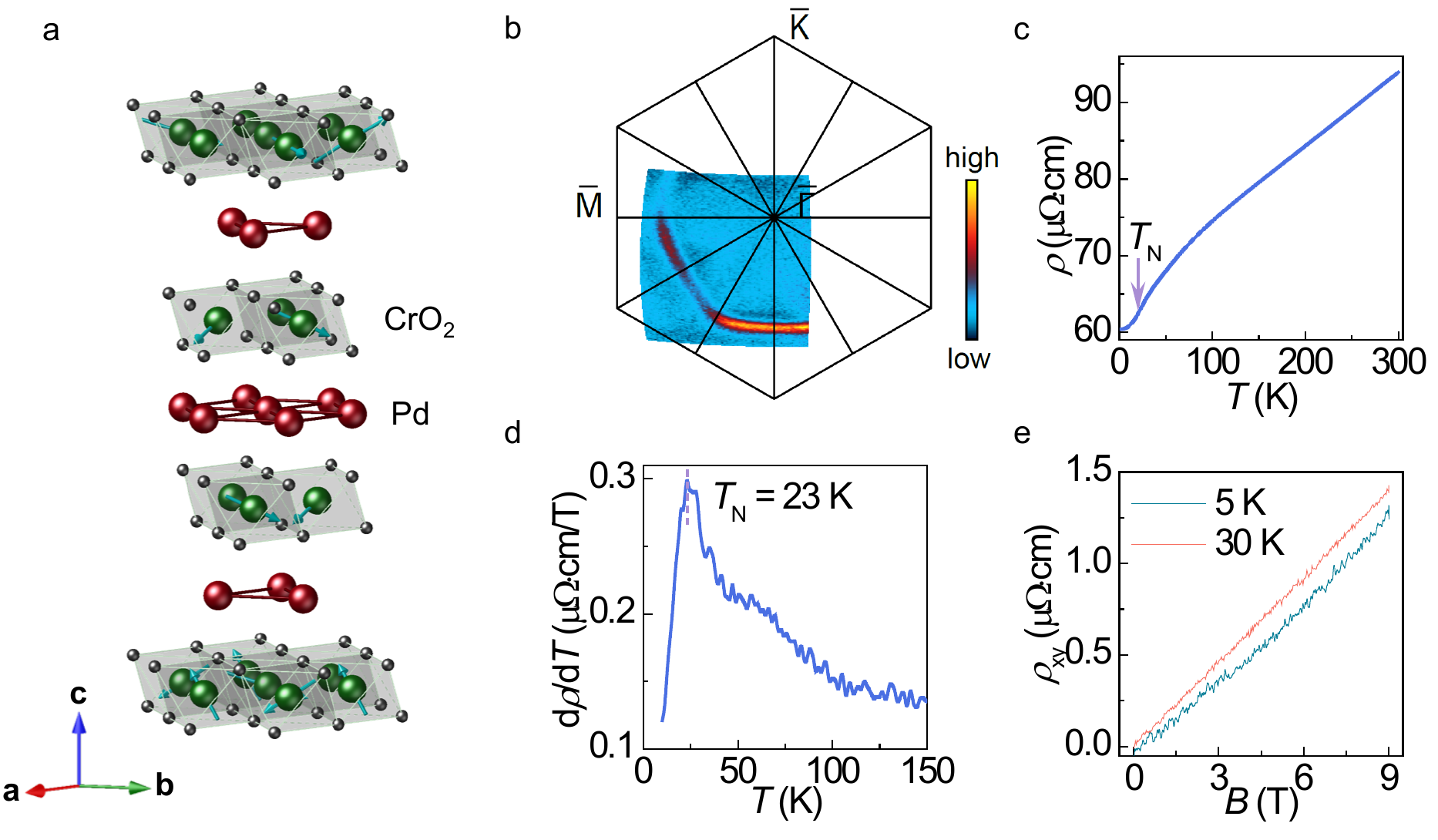}
    	\caption{\textbf{Structural, electronic and magnetic properties of PdCrO$_2$ samples.
        a, }Schematic depicting the crystal structure of PdCrO$_2$. The material consists of alternating atomic planes occupied of chromium atoms in green and palladium atoms in red separated by oxygen atoms in black. The arrows represent spin orientations in the Cr layers. \textbf{b, } Angle-resolved photoemission spectroscopy intensity maps of the Fermi surface of a $3$ nm PdCrO$_2$ sample collected using $21.2$ eV photons at $6$ K.
        \textbf{c, }Temperature-dependent resistivity of a $7$ nm thick PdCrO$_2$ sample.
        \textbf{d, } First derivative of resistance with respect to temperature for a $7$ nm thick PdCrO$_2$ film.  
        \textbf{e, } Hall response as a function of temperature for the $7$ nm thick PdCrO$_2$ film. A dc current of $100 \mu$A is applied and the magnetic field is swept out of the plane.}
    	\label{fig:1}
    	%\vspace{80pt}
    \end{figure}

    Here, we report measurements of the spin torque generated by the layered non-collinear antiferromagnet PdCrO$_2$ as a function of temperature. PdCrO$_2$ has the delafossite structure with space group $R \Bar{3} m$, consisting of alternating highly-conducting palladium layers and Mott-insulating CrO$_2$ layers within which the spins on the chromium sites form a 120$^\circ$ non-collinear structure (Figure~\ref{fig:1}a).\supercite{takatsu2009critical,ok2013quantum} The electrical conduction of PdCrO$_2$ is highly two dimensional and confined in the palladium layers, exhibiting a large anisotropy between conduction along the \textit{c} axis and within the basal plane.\supercite{takatsu2010anisotropy} The non-collinear spin structure within the chromium lattice is coupled to the highly-conducting palladium layers.\supercite{takatsu2014magnetic,takatsu2009critical,ghannadzadeh2017simultaneous,sunko2020probing} Novel physics including quantum oscillations,\supercite{ok2013quantum} loss of interlayer coherence,\supercite{ghannadzadeh2017simultaneous} unusual magnetothermopower,\supercite{arsenijevic2016anomalous} and an unconventional anomalous Hall effect\supercite{takatsu2010unconventional} have been observed in PdCrO$_2$ crystals.  
    %With advanced thin-film growth techniques, the strain, termination, and dimensions of PdCrO$_2$ thin films can now be precisely controlled by epitaxial growth.\supercite{song2024surface} 
    Because of the highly-anisotropic structure of PdCrO$_2$, with its insulating CrO$_2$ layers, one might expect vertically-flowing spin currents and associated spin torques to be highly suppressed, similarly to vertically-flowing charge currents. However, we find that this is not the case, and spin torque produced by PdCrO$_2$ is strongly enhanced upon cooling through the N\'eel temperature.
    \\
    
    %As illustrated in Fig.~\ref{fig:1}a, PdCrO$_2$ is formed by alternating Pd and Cr layers, with magnetic spins in Cr layers forming a 120$^\circ$ non-collinear structure. The spin-resolved energy bands of PdCrO$_2$ were calculated with first-principles density function theory (DFT) and are displayed in Fig.~\ref{fig:1}b. A clear spin splitting is observed when non-collinear antiferromagnetic ordering is introduced into the system, in good agreement with experimental observations \cite{sobota2013electronic, mazzola2018itinerant}.  The N\'eel temperature of a $7$ nm PdCrO$_2$ sample was determined to be approximately $23$ K by temperature dependent resistance (Fig.~\ref{fig:1}c), comparable to that of bulk PdCrO$_2$ ($37.5$ K) \cite{takatsu2010single,takatsu2010unconventional}. Another critical evidence to prove the existence of a broken TRS is anomalous Hall effect. Hall resistivity as a function of external magnetic field was shown for two temperatures-$5$ K and $30$ K. Below N\'eel temperature, a pronounced non-linear Hall response to external magnetic field is observed, agreeing with anomalous Hall effect seen in PdCrO$_2$ \cite{takatsu2010unconventional}, while the Hall resistivity scales linearly with external magnetic field in the paramagnetic state, indicating a preserved TRS. 
    \noindent 
    \textbf{Results}
    \\
    \noindent\textbf{Structural, electronic and magnetic properties of PdCrO$_2$ thin films}\\
    To measure the current-induced torques generated by PdCrO$_2$, we grew epitaxial thin films on (0001) sapphire substrates in a VEECO GEN10 MBE system as described in ref.\supercite{song2024surface} (Methods). %Figure \ref{fig:2}a shows an x-ray diffraction 2$\theta$-$\omega$ scan for a $5$ nm PdCrO$_2$ thin film which demonstrates single-phase growth. 
    %Details of the sample growth are provided in Methods. 
    The Fermi surface of a $3$ nm thick PdCrO$_2$ film as measured by angle-resolved photoemission spectroscopy (ARPES) at 6 K is shown in Figure~\ref{fig:1}b. A hexagonal pocket, centered at the $\bar{\Gamma}$ point, corresponds to the bulk state of PdCrO$_2$ contributed by electrons from the palladium layers. The metallic behavior of a 7 nm thick PdCrO$_2$ film presents an in-plane resistivity of 93 $\mu\Omega\cdot$cm at room temperature and 60.5 $\mu\Omega\cdot$cm at 2 K (Figure~\ref{fig:1}c). The N\'eel temperature of the $7$ nm thick PdCrO$_2$ sample was determined to be approximately $23$ K by temperature-dependent resistivity measurements (Figure~\ref{fig:1}d), reduced slightly relative to bulk PdCrO$_2$ ($37.5$ K).\supercite{takatsu2010single,takatsu2010unconventional}  (The N\'eel temperature for our PdCrO$_2$ films with various thicknesses ranged from 22 K to 30 K.\supercite{song2024surface}) The Hall resistivity as a function of an applied out-of-plane magnetic field has a nonlinear dependence at $5$ K and a simple linear dependence in the paramagnetic phase above the N\'eel temperature (Figure~\ref{fig:1}e), in agreement with previous measurements on PdCrO$_2$ single crystal.\supercite{takatsu2010unconventional}
    
    After deposition, our PdCrO$_2$ films were transferred through air to a separate sputter system, where they were pre-baked to 150 $^{\circ}$C to remove adsorbed water and cooled overnight.  A 5 nm thick Ni$_{80}$Fe$_{20}$ (Py) layer was then deposited by magnetron sputtering on top of the PdCrO$_2$ thin film (Methods), resulting in a magnetic layer with in-plane magnetic anisotropy.\\ %We measured no observable exchange coupling between the PdCrO$_2$ and the Py layer, in that the hysteresis curves as a function of in-plane magnetic field for the Py resistance were symmetric about zero applied magnetic field  (Fig.~\ref{fig:2}b).

\begin{figure}[!hb]
    	\centering
        \includegraphics[width=1\textwidth]{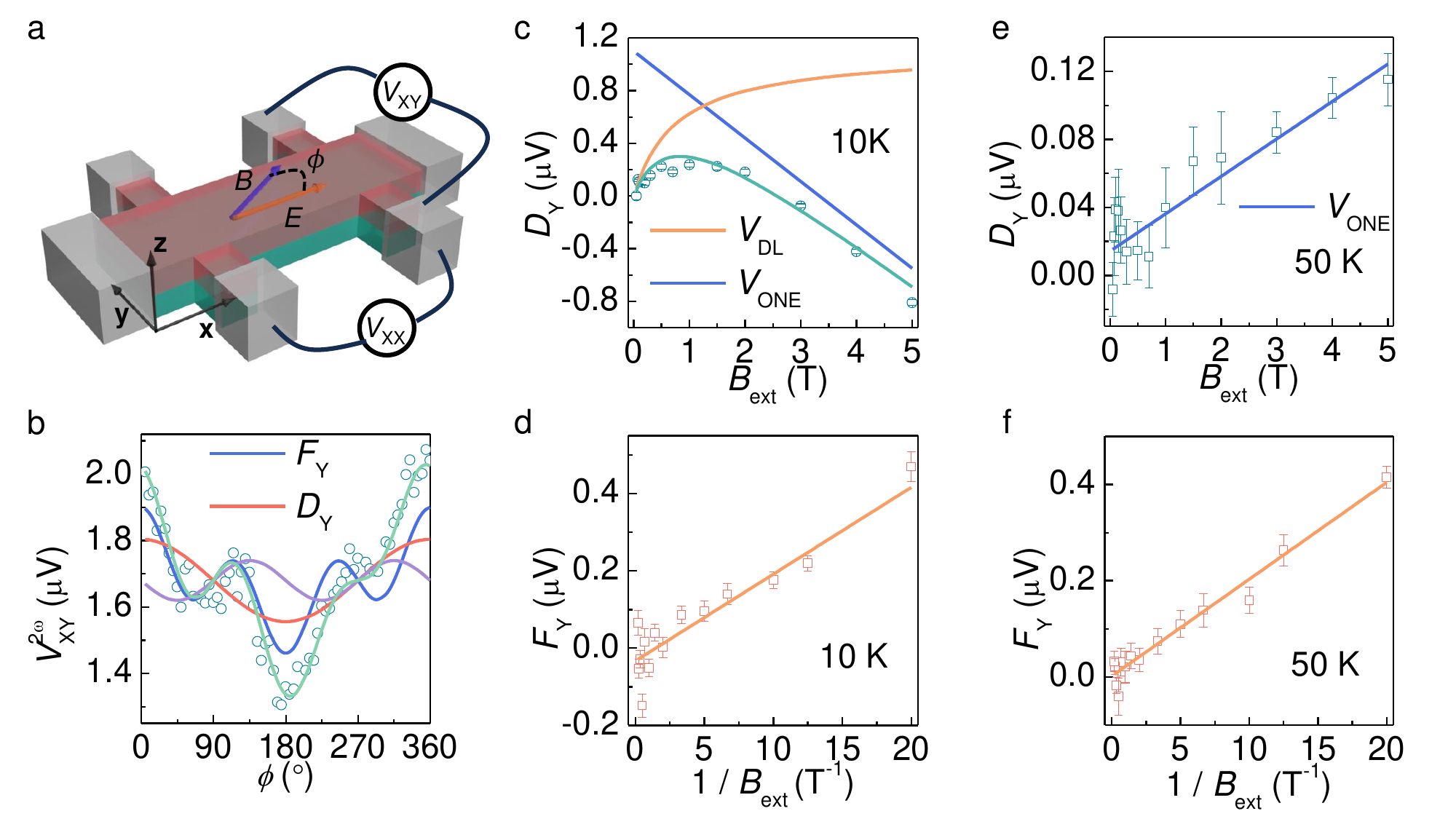}
    	\caption{\textbf{Second harmonic Hall measurements on PdCrO$_2$ ($7$ nm) / Py ($5$ nm).
        a, }Schematic for the Hall devices containing a PdCrO$_2$ / Py bilayer. For second-harmonic Hall measurements, the external magnetic field is applied in plane at an angle $\phi$ relative to the applied electric field $E$. The PdCrO$_2$ and Py layers are shown in teal and red, respectively.
        \textbf{b, }Second harmonic Hall voltage as a function of $\phi$ at $10$ K. The strength of the magnetic field is $800$ Oe. The $\cos\phi$ ($D_Y$) and $\cos\phi \cos2\phi$ ($F_Y$) components are shown in red and blue, respectively. A $sin2\phi$ component caused is shown in purple.
        \textbf{c, }Magnetic-field dependence of the $\cos\phi$ component of the second harmonic Hall voltage ($V_{XY}^{2\omega}$) at $10$ K. The solid curve in teal is a sum of voltages from damping-like torque (orange), the ordinary Nernst effect (blue) and the spin Seebeck effect (a constant term).
        \textbf{d, }Magnetic field dependence of the $\cos\phi \cos2\phi$ contribution to the second harmonic Hall voltage as a function of $1/B_{ext}$ at $10$ K.
        \textbf{e, }Magnetic field dependence of the $\cos\phi$ component of the second harmonic Hall voltage ($V_{XY}^{2\omega}$) at $50$ K. The solid curve in teal is a linear fit to the $D_Y$ component at $50$ K.  
        \textbf{f, }Magnetic field dependence of the $\cos\phi \cos2\phi$ contribution to the second harmonic Hall voltage as a function of $1/B_{ext}$ at $50$ K.}
    	\label{fig:2}
    	%\vspace{80pt}
    \end{figure}

    \noindent
    \textbf{Second harmonic Hall voltage measurement}\\
    The current-induced spin torque was characterized by two separate techniques: second-harmonic Hall measurements\supercite{avci2014interplay,hayashi2014quantitative} and spin-torque ferromagnetic resonance (ST-FMR).\supercite{liu2011spin, liu2012spin} The second-harmonic Hall measurements were performed using the Hall-bar geometry shown in Figure~\ref{fig:2}a, with a channel width of 20 $\mu$m, and Hall-probe widths of 5 $\mu$m (Methods).  %{\color{blue}[A note here for future reference: For measuring 1st harmonic Hall signals, the width of the Hall probes relative to the channel width does not matter, but this is not the case for 2nd harmonic Hall signals. To avoid a reduction of the 2nd harmonic signal due to current spreading at the junction, you should make the Hall probes narrow compared to the channel.]\color{red}[The Hall probe width is 5 $\mu$m and 30 $\mu$m is the distance between the two adjacent Hall probes.]}  %Example second harmonic Hall data for a $5$ nm PdCrO$_2$/5 nm Py sample and the resulting analysis are shown in Fig.~\ref{fig:3}.  
    In response to an oscillating applied current with amplitude $I_0$ and frequency $\omega$ ($I_0$ = $10$ mA (peak) and $\omega$ = $17$ Hz in our case), current-induced torques cause the magnetization orientation to oscillate, and consequently the anomalous and planar Hall resistances also oscillate with frequency $\omega$. Mixing between the oscillating current and the oscillating Hall resistances causes the Hall voltage to oscillate at frequency $2\omega$. Since (0001)-oriented PdCrO$_2$ is a high-symmetry material and forms structural twin domains\supercite{harada2018highly,song2024surface}, only torques with conventional symmetries are allowed.  The damping-like and field-like torques therefore generate second-harmonic Hall signals of the form\supercite{hayashi2014quantitative,avci2014interplay,macneill2017,wen2017temperature,gupta2020manipulation,bose2022tilted}
    \begin{equation}
    \label{V_XY}
        V_{XY}^{2\omega}=D_Y \cos(\phi)+F_Y \cos(\phi)\cos(2\phi)+C,
    \end{equation} 
    where the coefficient $D_Y$ contains the contribution from the damping-like component of torque and $F_Y$  contains the contribution from the field-like component. An example second-harmonic Hall signal for a PdCrO$_2$ (7 nm) / Py (5 nm) sample at $10$ K is shown in Figure~\ref{fig:2}b, along with a fit to Equation~\ref{V_XY}.  The quality of the fit is improved by adding a small additional $\sin(2\phi)$ component (purple), which is likely from the planar Nernst effect due to an in-plane thermal gradient.\supercite{avery2012observation, avery2012determining,karimeddiny2020transverse}
    %The fit is not improved with the inclusion of additional terms corresponding to unconventional current-induced torques.

    The coefficient $D_Y$ contains contributions from not only the damping-like torque but also thermoelectric effects (for example, spin Seebeck, anomalous Nernst, and ordinary Nernst effects), and similarly the coefficient $F_Y$ contains a contribution from the current-generated Oersted field as well as the field-like torque. The damping-like torque term can be isolated based on different dependencies on the strength of the applied magnetic field, $B_{ext}$,\supercite{gupta2020manipulation,bose2022tilted} and the Oersted contribution can be calculated based on the measured current density. We extract the effective fields corresponding to the damping-like and field-like torques ($\Delta B_{DL}$ and $\Delta B_{FL}$) according to
    \begin{equation}
    \label{eqn:D_Y}
        D_Y= \frac{I_0 R_{AHE}}{2}\frac{\Delta B_{DL}}{B_{ext}+\mu_0M_{eff}}+V_{ONE}B_{ext}+V_{SSE}
    \end{equation}
    
    \begin{equation}
    \label{eqn:F_Y}
        F_Y= I_0 R_{PHE} \frac{\Delta B_{FL}}{B_{ext}},
    \end{equation}
    where $\mu_0M_{eff}$ is the effective anisotropy field, $\Delta B_{FL}$ contains contributions from both the field-like spin-orbit torque and the Oersted field from the applied current, $V_{ONE}$ is the voltage from the ordinary Nernst effect, and $V_{SSE}$ is the voltage from the spin Seebeck effect. The anomalous Hall resistance coefficient $R_{AHE}$ was measured by sweeping a magnetic field from $3$ T to $-3$ T in the out-of-plane direction, and the planar-Hall coefficient $R_{PHE}$ was extracted from measurements as a function of in-plane magnetic field angle. Example temperature dependences of $R_{AHE}$ and $R_{PHE}$ are summarized in Supplementary Information Section 3.

    Figure~\ref{fig:2}c and Figure~\ref{fig:2}d show the values of $D_Y$ and $F_Y$  measured at $10$ K for the PdCrO$_2$ (7 nm) / Py (5 nm) sample with fits to Equation~\ref{eqn:D_Y} and Equation~\ref{eqn:F_Y}. The dependence of $D_Y$ on $B_{ext}$ in Figure~\ref{fig:2}c is nonlinear, indicating the presence of a non-zero damping-like torque.  %From the fits, we get the damping-like and field-like effective field per unit current density $\Delta B_{DL}/J_{AF} = (0.88\pm 0.10)\times 10^{-13}$ T per A/m$^2$, $(\Delta B_{FL} + B_{Oe})/J_{AF} = (0.31\pm0.02)\times 10^{-14}$ T per A/m$^2$ and $V_{ONE} = -0.17 \pm 0.02$ $\mu$V, $V_{SSE} = 0.093 \pm 0.061$  $\mu$V. 
    Similar fits for $D_Y$ and $F_Y$ at $50$ K, above the N\'eel temperature of the $7$ nm thick PdCrO$_2$ film, are shown in Figure~\ref{fig:2}e and Figure~\ref{fig:2}f. At this temperature, the dependence of $D_Y$ on $B_{ext}$ is to a good approximation linear, indicating that the damping-like torque is close to zero. The dependence of the field-like component $F_Y$ at $50$ K (Figure~\ref{fig:2}f) remains similar to $F_Y$ at $10$ K (Figure~\ref{fig:2}d), indicating that the field-like torque has negligible dependence on the antiferromagnetic ordering.  %suggesting that the Oersted field is the dominant contribution to $F_Y$ in both cases. %Representative second harmonic Hall measurements for the pertinent samples presented in this study are displayed in Supplemental Information Section 2. 
    The damping-like spin-orbit torque efficiency from second harmonic Hall measurements can be estimated with
    \begin{equation}
    \label{xi_SHH}
    \xi_{DL}^Y = \frac{2e}{\hbar}\frac{\Delta B_{DL}}{J_{AF}}M_s t_{Py},
    \end{equation}
    where $M_s$ is the saturation magnetization of Py, $t_{Py}$ is the thickness of the Py, and we estimate the current density in the PdCrO$_2$ ($J_{AF}$) from a parallel-resistor model based on the relative resistivities of PdCrO$_2$ and Py (see Supplementary Information Section 4).
    
\begin{figure}[!hb]
    	\centering
        \includegraphics[width=0.8\textwidth]{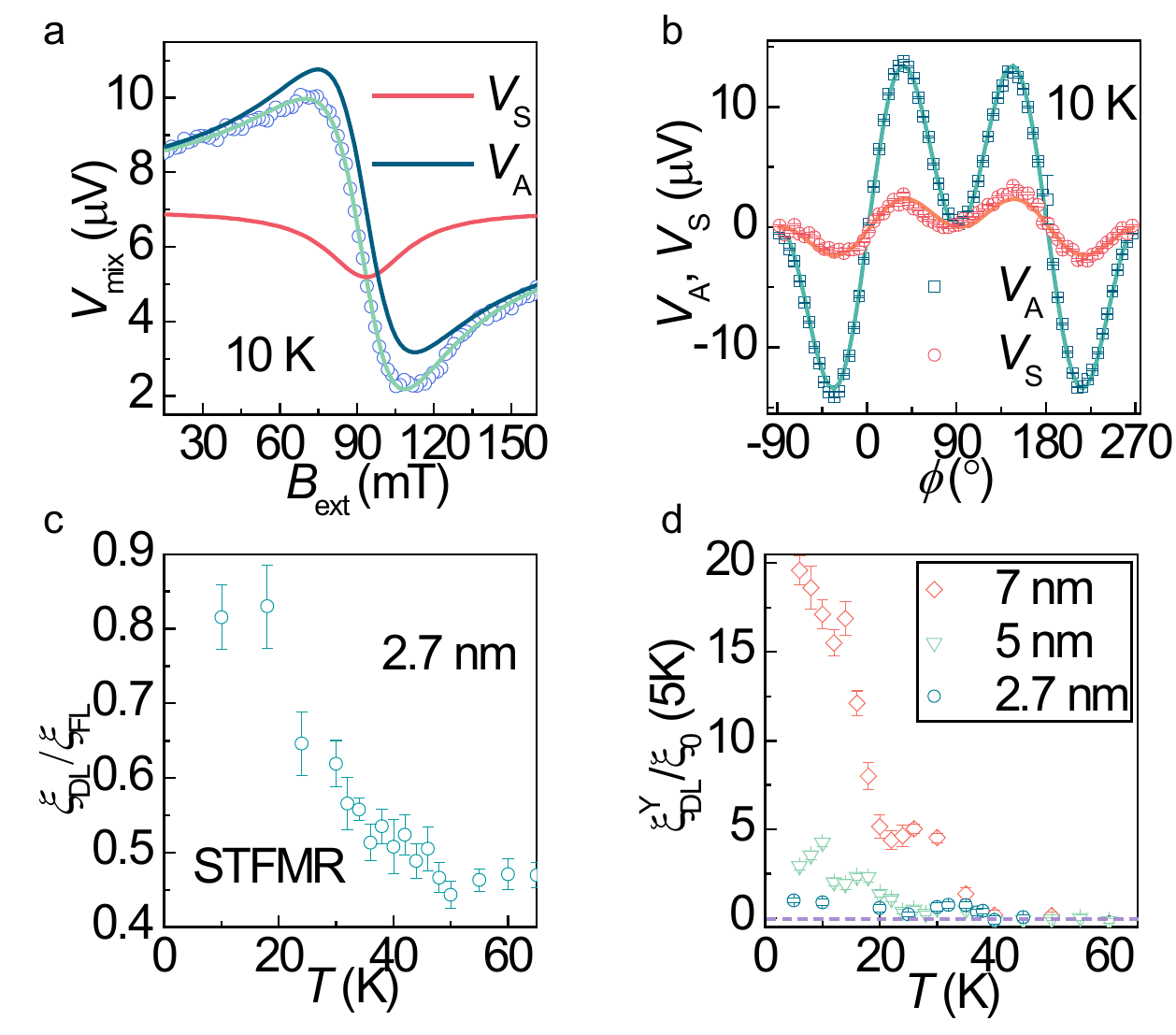}
    	\caption{\textbf{Spin-orbit-torque generation in PdCrO$_2$ ($2.7$ nm) / Py ($5$ nm) thin film measured with longitudinal ST-FMR.
        a, }An example ST-FMR scan at $10$ K, where the longitudinal mixing voltage ($V_{mix}$) is plotted against external magnetic field ($H_{ext}$). Symmetric and anti-symmetric components are represented by red and blue curves, respectively. The microwave current applied has a frequency of $9$ GHz and a power of $18$ dBm. The external magnetic field is oriented at $45^\circ$ to the current. 
        \textbf{b, }Anti-symmetric mixing voltage ($V_A$) and symmetric mixing voltage ($V_S$) as a function of magnetic field angle ($\phi$). Blue and red curves are fits to $\sin2\phi \cos\phi$. 
        \textbf{c, }In plane damping-like spin-orbit-torque efficiency ratio ($\xi_{DL}/\xi_{FL}$) from STFMR as a function of temperature.
        \textbf{d, }Spin-orbit-torque efficiency from second harmonic Hall measurements as a function of temperature for PdCrO$_2$ of various thicknesses. %Only relative values of spin Hall conductivities are shown for comparison.
        } 
    	\label{fig:3}
    	%\vspace{80pt}
    \end{figure}

    From the results of similar second harmonic Hall measurements, the full temperature dependence of the spin-orbit torque efficiency corresponding to the anti-damping torque is plotted in Figure~\ref{fig:3}d for PdCrO$_2$ films with thicknesses of 2.7 nm, 5 nm, and 7 nm.
    In this graph we plot only relative values (i.e., we normalize relative to the spin-orbit torque efficiency of 2.7 nm PdCrO$_2$ at 5 K) because the fitted value of $\xi^Y_{DL}$ can vary significantly depending on the choice of the value for $\mu_0M_{eff}$, and in some cases the extracted value of $\xi^Y_{DL}$ can appear to be much larger than the value indicated by ST-FMR (see below).  Nevertheless, it is clear from these second-harmonic Hall measurements that the anti-damping current-induced torque turns on dramatically as the temperature is cooled below 25-30 K, the N\'eel temperature of the PdCrO$_2$.   The spin-orbit torque efficiency also increases as a function of PdCrO$_2$ thickness as summarized in Figure~\ref{fig:3}d, which suggests a bulk origin for the spin current generation with a spin diffusion length of order a few nanometers, rather than an interface mechanism.\\

\begin{figure}[!hb]
    	\centering
        \includegraphics[width=1\textwidth]{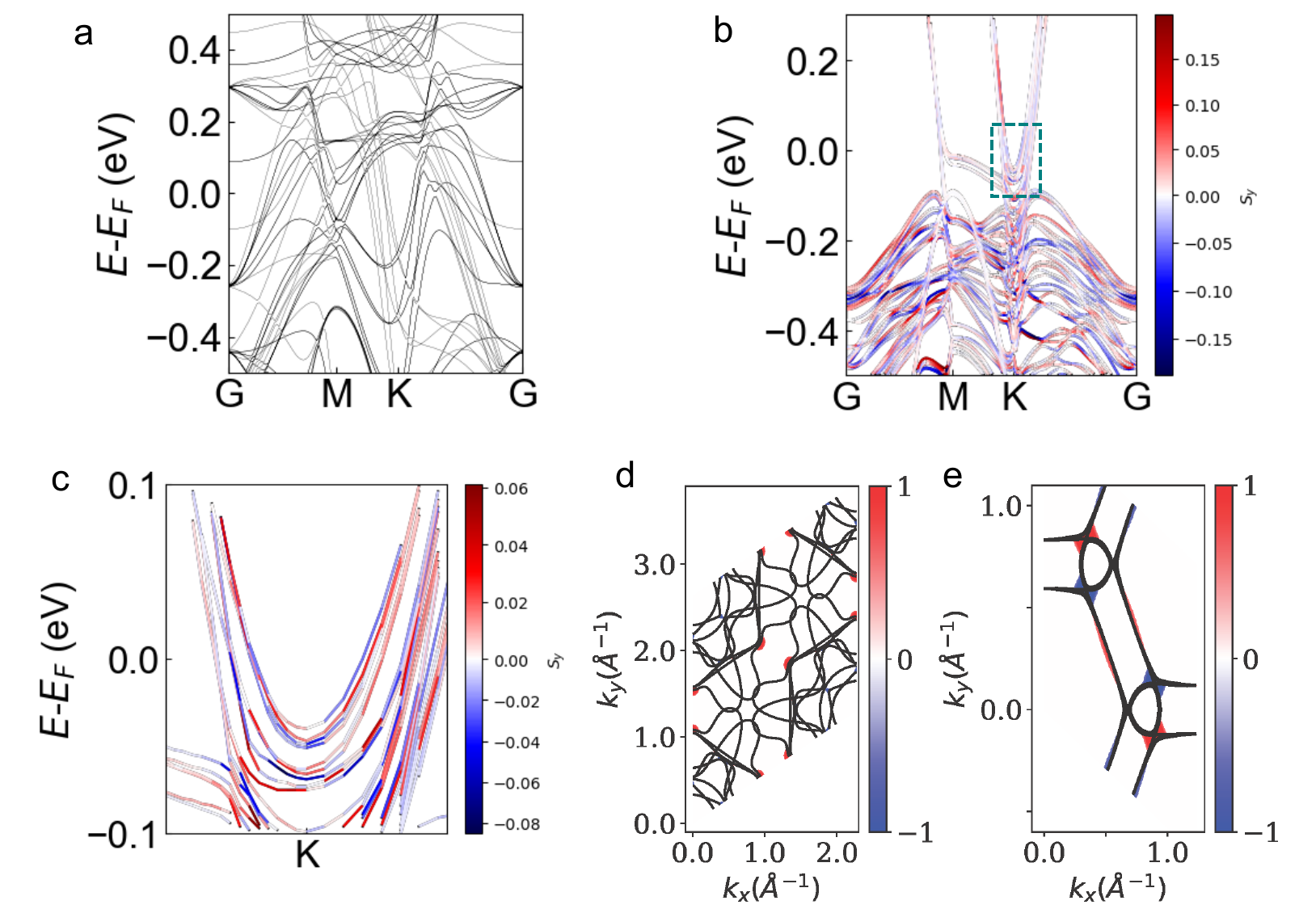}
    	\caption{\textbf{Electronic structure for PdCrO$_2$.
        a, }Electronic band structure of paramagnetic PdCrO$_2$.
        \textbf{b, } Spin-resolved electronic band structure of antiferromagnetic PdCrO$_2$. Red and blue represent positive and negative spin orientations, respectively. \textbf{c, }Electronic band structure of the enclosed rectangular area in \textbf{b}. \textbf{d, }Spin Berry curvature distribution in the plane formed by $k_x$ and $k_y$ at $k_z$ = 0 for the paramagnetic phase. \textbf{e, } Spin Berry curvature distribution in the plane formed by $k_x$ and $k_y$ at $k_z$ = 0 for the antiferromagnetic phase.
        }
    	\label{fig:4}
    	%\vspace{80pt}
    \end{figure}

    \noindent
    \textbf{Spin-torque ferromagnetic resonance measurement}\\
    ST-FMR measurements were performed to confirm the results of the second-harmonic Hall measurements. The ST-FMR device consists of a PdCrO$_2$ ($2.7$ nm) / Py ($5$ nm) bilayer. (We choose to use a relatively thin PdCrO$_2$ layer ($2.7$ nm) in order to minimize interference from artifact voltages induced by spin pumping\supercite{cham2022separation,liu2021influence,dong2023enhancement} even though this reduces the value of the measured spin Hall ratio.) An external magnetic field is oriented at an angle $\phi$ to the current direction. The magnetoresistance that oscillates at the same frequency as the input current mixes with the current and produces a dc mixing voltage that can be expressed as \supercite{liu2011spin,liu2012spin,mellnik2014spin,karimeddiny2020transverse}
    \begin{equation}
    \label{Vmix}
    V_{mix} = \frac{I_{rf}R_{AMR}}{2\alpha \omega^+} \sin2\phi \cos\phi \times (F_S(H_{ext})\tau_{DL}^0 + F_A(H_{ext})\frac{\omega_2}{\omega}\tau_z^0),
    \end{equation}
    where $I_{rf}$ is the total microwave current flowing into the device, $R_{AMR}$ is the anisotropic magnetoresistance, $\tau^0_{DL}$ and $\tau^0_z$ are the current-induced damping-like torque and field-like torque, respectively, $F_S(H_{ext}) = \Delta^2/[\Delta^2 + (H_{ext}-H_0)^2]$ is the symmetric Lorentzian function, $F_A(H_{ext}) = F_S(H_{ext})(H_{ext}-H_0)/\Delta$ is the anti-symmetric Lorentzian function,  $\Delta = \alpha \omega / \gamma$ is the half-width-at-half-maximum linewidth, $\omega_1=\gamma H_0$, $\omega_2=\gamma(H_0+ M_{eff})$, $\omega^+=\omega_1+\omega_2$, and $\omega = \sqrt{\omega_1 \omega_2}$ at resonance. Figure \ref{fig:3}a displays an example ST-FMR spectrum for $I_{rf}$ at $9$ GHz and $18$ dBm, with a fit to a sum of symmetric and anti-symmetric Lorentzian functions. The  ST-FMR scans were performed with the external magnetic field oriented at various angles to the current direction (Figure~\ref{fig:3}b).  Both the symmetric ($V_S$) and anti-symmetric ($V_A$) voltages are well described by the angular dependence $\sin2\phi \cos\phi$ associated with current-induced torques having conventional symmetry. %The current-induced torques in the form of $m\times(\sigma \times m)$ or $\sigma \times m$ have a $\cos\phi$ angular dependence. The mixing of the current-induced torque with the magnetoresistance ($dR(\phi)/d\phi \sim \sin2\phi$) gives rise to the same angular dependence for $V_S$ and $V_A$: $\sin2\phi \cos\phi$. 
Based on the second harmonic Hall results, the field-like torque has little temperature dependence near the N\'eel temperature of the PdCrO$_2$, so to illustrate the temperature dependence of the anti-damping spin-orbit-torque efficiency we plot the ratio\supercite{liu2011spin,karimeddiny2020transverse} 
    \begin{equation}
    \label{xiratio}
        \frac{\xi_{DL}}{\xi_{FL}}= \frac{V_S}{V_A} \sqrt{1+\frac{M_{eff}}{H_0}}, 
    \end{equation}
%    \begin{equation}
%    \label{xi}
%        \xi^Y_{DL} = \frac{V_S}{V_A} \frac{e\mu_0 M_S t_{FM}t_{AF}}{\hbar}[1+ M_{eff}/H_0]^{1/2}
%    \end{equation}
% where $e$ is the electronic charge, $M_S$ is the saturation magnetization of the ferromagnet, and $\hbar$ is Planck's constant.  
The temperature dependence of this ratio based on the ST-FMR measurements is shown in Figure~\ref{fig:3}c with a comparison to the temperature dependence determined from the second harmonic Hall measurements in Figure~\ref{fig:3}d. 
The ST-FMR measurements confirm the primary finding of this paper: that the damping-like spin torque increases dramatically upon reducing the temperature
through the N\'eel temperature of PdCrO$_2$.  We therefore conclude that the current-induced torque in this case is dominated by a mechanism dependent on the magnetic ordering.  If we assume that the field-like torque is due entirely to the Oersted field, then the ST-FMR measurements for the PdCrO$_2$ ($2.7$ nm) / Py ($5$ nm) yield a spin Hall ratio of $\xi_{SH} = 0.027$ and a spin Hall conductivity of $\sigma_{SH} = 6.8 \times 10^3$ ($\hbar$/2e)$\Omega^{-1}$m$^{-1}$, comparable to that reported in non-collinear antiferromagnet Mn$_3$Ir\supercite{zhang2016giant}.  This is an underestimate for the maximum anti-damping torque efficiency that can be produced by PdCrO$_2$ both because there appears to be a field-like spin-orbit torque with the same sign as the Oersted torque (see Supplementary Information Section 4), and because we expect larger values for samples with a PdCrO$_2$ layer thickness greater than the spin diffusion length. (We have not performed those measurements.) For clarity, we repeat that for (0001)-oriented PdCrO$_2$, the anti-damping torque is constrained by symmetry to be oriented only in-plane and perpendicular to the charge current, which is in contrast to the non-collinear antiferromagnet Mn$_3$Sn in which an out-of-plane spin component is allowed and observed.\supercite{kimata2019magnetic,hu2022efficient}
\\
   \noindent
   \textbf{Electronic band structure of PdCrO$_2$}\\
    These experimental results can be compared to predictions of density functional theory for PdCrO$_2$ in the paramagnetic and antiferromagnetic states (Figure~\ref{fig:4}). The paramagnetic phase has doubly-degenerate bands throughout the Brillouin Zone (Figure~\ref{fig:4}a) and the only source of the intrinsic contribution to the spin Hall conductivity is due to the closely spaced bands around the Fermi levels. For the non-collinear antiferromagnetic phase, although there are fewer closely spaced bands around the Fermi level, along certain directions (e.g., around $K$ point as shown in Figure~\ref{fig:4}b, c) the spin resolved band structure has a spin-split behavior analogous to the structure of ferromagnets and altermagnets.\supercite{yuan2020giant,vsmejkal2022emerging} This shows that in addition to the spin Hall conductivity, a non-zero magnetic spin Hall conductivity is also possible.  Figure~\ref{fig:4}d and Figure~\ref{fig:4}e show the heatmap of the spin Berry curvature within the $k_z=0$ plane in the nonmagnetic and non-collinear antiferromagnetic phases. The Fermi contours for the nonmagnetic phase show that there are many bands closely spaced in the nonmagnetic phase but only few regions show significant hot spots contributing to the spin Hall conductivity. The non-collinear antiferromagnetic phase having fewer closely spaced spin-split bands possesses regions with significantly larger spin Berry curvature.  The predicted intrinsic spin Hall conductivities are $8.2\times10^3$ ($\hbar$/2e)$\Omega^{-1}\cdot$m$^{-1}$ for the non-collinear antiferromagnetic phase and $3.3\times10^3$ ($\hbar$/2e)$\Omega^{-1}\cdot$m$^{-1}$ in the non-magnetic phase. The predicted variation in the intrinsic spin Hall conductivity as a function of varying the Fermi energy is shown in Figure\ S6a in the supplemental information.  We also show in Figure\ S6b a calculation for the magnitude of the extrinsic magnetic spin Hall conductivity arising from the spin-split bands.  The strength of this contribution depends on what value is assumed for the scattering rate, but it is predicted to be comparable to or possibly even greater than the intrinsic spin Hall conductivity for the antiferromagnetic state. In addition, in the presence of antiferromagnetic order there may be extrinsic contributions due to scattering from magnetic fluctuations that we do not attempt to calculate.\\  

\noindent
\textbf{Conclusion}\\
In summary, we have demonstrated that the non-collinear antiferromagnet PdCrO$_2$ can  produce current-induced damping-like torque acting on an adjacent ferromagnetic Permalloy layer. The sharp onset of spin current generation associated with antiferromagnetic transition indicates a mechanism dependent on magnetic ordering for spin current produced in PdCrO$_2$. Furthermore, the much weaker torque produced above N\'eel temperature indicates that spin-orbit coupling alone, although present in the system, does not contribute significantly to spin current generation. 
\clearpage  
%\printbibliography[title={References and Notes}]
\noindent 
\textbf{Materials and Methods}
\\
\noindent\textbf{Sample growth and device fabrication}\\
PdCrO$_2$ films were prepared using molecular-beam epitaxy (MBE) in a Veeco GEN10 MBE system on (0001) sapphire substrates. The substrates were heated to 600 $^\circ$C during deposition, with temperature monitored by an optical pyrometer operating at a wavelength of 980~nm. The deposition atmosphere consisted of a background pressure of 5 × 10$^{-6}$ Torr with a gas mixture of 80\% ozone and 20\% oxygen. Achieving single-phase PdCrO$_2$ films requires the use of a delafossite-structured buffer layer on the (0001) sapphire substrates. In this work, a 12-nm-thick insulating CuCrO$_2$ film was employed to stabilize the PdCrO$_2$ phase.  Additional details about the PdCrO$_2$ growth process can be found in the supplementary material of Ref.\supercite{song2024surface}.
After the growth of the PdCrO$_2$, the samples were transferred through air to a sputtering system with a base pressure of less than $3\times 10^{-8}$ Torr.  Prior to the deposition of the Py layer, the PdCrO$_2$  films were pre-baked at 150$^{\circ}$C for 10 minutes in vacuum in the sputtering chamber and were cooled overnight to remove adsorbed water and other possible contaminants on the sample surface. Subsequently, $5$ nm of Py (Ni$_{80}$Fe$_{20}$) was grown and capped with $1.5$ nm of Ta to prevent oxidation. Hall bars of dimension $20 \times 95$ $\mu$m$^2$ were patterned with photolithography and ion milling. Then Ti/Pt contacts were patterned with an aligned exposure, and deposited with sputtering and lift-off. 
\vspace{5mm}
\\
\noindent\textbf{Second harmonic Hall measurements}\\
The second harmonic Hall measurements were performed in a Quantum Design EverCool PPMS with a maximum magnetic field of $9$ T. The samples were rotated in the plane with a rotator. The samples were mounted on the rotator such that magnetic field was parallel to the current direction at $0^\circ$. The ac current of $17$ Hz was applied using the internal oscillator of a Signal Recovery DSP 7265 lock-in amplifier. A Keithley 6221 ac source meter was used as the ac current source when Stanford Research SR 830 lock-in amplifiers were used. The first and second harmonic voltages were detected with the same Signal Recovery DSP 7265 lock-in amplifier or SR 830 lock-in amplifiers.  
\vspace{5mm}
\\
\noindent\textbf{ DFT Computational Details}\\
DFT calculations are performed using the Vienna \textit{ab initio} simulation package (VASP).\supercite{vaspcalc} We apply generalized gradient approximation (GGA) using the projector augmented-wave method. In the calculations, we use nonmagnetic PdCrO$_2$ in $R\overline{3}m$ space group and $\sqrt{3}\times\sqrt{3}\times 2$ supercell is used for the antiferromagnetic phase.  We have used the model 4 antiferromagnetic phase in the reference\supercite{takatsu2014magnetic} as our starting point for the calculations.  The atomic structures in antiferromagnetic phase are fully relaxed using the $6\times6\times2 \quad \Gamma-$ centered k-point grid and kinetic energy cutoff of $450$ eV. The relaxation is performed until the atomic forces on each atom are converged to better than 0.01 $eV/\AA$. Spin-orbit coupling is included in all calculations.

In the diffusive-transport regime, the spin conductivity has two contributions:

\begin{equation}
\label{oddshc}
    \sigma^k_{ij}=-\frac{e\hbar}{\pi} \int \frac{d^3k}{(2\pi)^3} \sum_{n,m} \frac{\Gamma^2 Re(\langle n \vec{k} |J^k_i |m \vec{k}\rangle \langle m \vec{k} |v_j |n \vec{k}\rangle )}{[(E_F-E_{n \vec{k}})^2+\Gamma^2][(E_F-E_{m \vec{k}})^2+\Gamma^2]},
\end{equation}

\begin{equation}
\label{evenshc}
\begin{split}
    \sigma_{ij}^k=&-\frac{2e^2}{\hbar} \int \frac{d^3k} {(2\pi)^3} \sum_n f_{n \vec{k}}  Im \sum_{n \neq m} \frac{\langle n \vec{k} | J^k_i | m \vec{k} \rangle  \langle m \vec{k} | v_j | n \vec{k} \rangle}{(E_{n \vec{k}}-E_{m \vec{k}})^2}\\
    =&\frac{e^2}{\hbar} \int \frac{d^3k} {(2\pi)^3} \sum_n f_{n \vec{k}}\Omega^k_{n,ij} (\vec{k})
\end{split}
\end{equation}
where $f_n(\vec{k})$ is the Fermi-Dirac distribution for the nth band, $J_i^k=\frac{1}{2}{v_i,s_k}$ is the spin current operator with spin operator $s_k, v_j=\frac{1}{\hbar} \frac{\partial H}{\partial k_j}$ is the velocity operator, and $i,j,k = x,y,z$. $\Omega_{n,ij}^k (\vec{k})$ is referred to as the spin Berry curvature in analogy to the ordinary Berry curvature. The spin conductivity given by \eqref{oddshc} is the Fermi surface property odd under time reversal symmetry ($\hat{T}$-odd) and \eqref{evenshc} is interband contributions which are even under time reversal symmetry ($\hat{T}$-even) .Spin conductivities were calculated using a Wannier linear response code\supercite{wanresponse} based on the tight-binding Hamiltonian utilizing the maximally localized Wannier functions obtained using a Wannier90 code.\supercite{marzari2012maximally} The spin Hall conductivities (given by Equation \eqref{evenshc}) were calculated using the tight-binding Hamiltonians with a $200 \times 200\times 2$ k-point mesh to get the converged values.
\\
\clearpage
\printbibliography[title={References}]
\clearpage
\noindent\textbf{Acknowledgments}
\\
\noindent
X.H.\ led the research, supported by the US Department of Energy, DE-SC0017671. Q.S.\ acknowledges the support from the U.S. Department
of Energy, DE-SC0002334 and Gordon and Betty Moore Foundation’s EPiQS Initiative (Grant Nos. GBMF3850 and GBMF9073 to Cornell
University). D.A.P.\ and D.G.S.\ acknowledge support from SUPREME, one of seven centers in JUMP 2.0, a Semiconductor Research Corporation(SRC) program sponsored by DARPA; R.J.\ and M.O.\ acknowledge support by the National Science Foundation (DMR-2104268) and R.J.\ thanks the support of a Kavli Institute at Cornell Graduate Engineering Fellowship; T.M.C. acknowledges support by the AFOSR/MURI project 2DMagic (FA9550-19-1-0390).  The research at University of Nebraska-Lincoln was supported by the Office of Naval Research (ONR grant N00014-20-1-2844) and by the National Science Foundation through EPSCoR RII Track-1 program (NSF Award OIA-2044049). The devices were fabricated using the shared facilities of the Cornell NanoScale Facility, a member of the National Nanotechnology Coordinated Infrastructure (supported by the NSF via grant NNCI-2025233) and the facilities of the Cornell Center for Materials Research.
\vspace{5mm}
\\
\noindent\textbf{Author contributions}\\
X.H. and D.C.R. conceived the research and led the mentoring of several junior students who contributed. Q.S. carried out the oxide thin film growth and performed X-ray diffraction and angle-resolved photoemission spectroscopy measurements with advice from D.G.S.. X.H. and M.O. performed the ferromagnet sputtering. X.H., D.A.P., and Y.C. conducted magnetization characterization. X.H. and D.A.P. performed device fabrication. X.H., D.A.P., and R.J. performed the second harmonic Hall measurements. X.H., D.A.P., and T.M.C. conducted the ST-FMR measurements.  X.H. and D.A.P. performed the data analysis and curation. G.G. performed the first-principles density functional theory calculations. X.H., D.A.P., and D.C.R. wrote the manuscript. All authors discussed the results and commented on the manuscript. 
\vspace{5mm}
\\
\noindent\textbf{Competing interests}\\
The authors declare no competing interests.
\vspace{5mm}
\\
\noindent\textbf{Data and materials availability}\\ The data that support the ﬁndings of this study are available from the corresponding author upon reasonable request.
     
\clearpage
\end{document}